\documentclass[a4paper,11pt]{article}
\pdfoutput=1 

\usepackage{jcappub} 

\usepackage[T1]{fontenc} 
\usepackage{amssymb}
\usepackage{amsmath}

\usepackage{todonotes}
\usepackage{graphicx,setspace}
\usepackage{dcolumn}
\usepackage{bm}
\usepackage[mathlines]{lineno}
\usepackage{changes}

\newcommand{\be}{\begin{equation}}
\newcommand{\ee}{\end{equation}}
\newcommand{\bea}{\setlength\arraycolsep{2pt} \begin{eqnarray}}
\newcommand{\eea}{\end{eqnarray}}

\title{Shadow cast and Deflection angle of Kerr-Newman-Kasuya spacetime}

\author[a,b,1]{Ali \"{O}vg\"{u}n,\note{Corresponding author.}}
\author[b]{\.{I}zzet Sakall{\i},}
\author[a,2]{Joel Saavedra}

\affiliation[a]{Instituto de F\'{\i}sica, Pontificia Universidad Cat\'olica de Valpara\'{\i}%
so, Casilla 4950, Valpara\'{\i}so, Chile.}
\affiliation[b]{Physics Department, Arts and Sciences Faculty, Eastern Mediterranean
University, Famagusta, North Cyprus via Mersin 10, Turkey.}

\emailAdd{ali.ovgun@pucv.cl} 
\emailAdd{izzet.sakalli@emu.edu.tr}
\emailAdd{joel.saavedra@ucv.cl}

\abstract{
We study the shadow cast or silhouette generated by a Kerr-Newman-Kasuya (KNK) spacetime (rotating dyon black hole). It is shown that in addition to the angular momentum of the black hole, the dyon charge also affects the
shadow image of the KNK black hole. Moreover, we analyze the weak gravitational
lensing by the KNK black hole by using the Gauss-Bonnet theorem. Finally, we find
that extra dyon charge decreases both the deflection angle and shadow of the
KNK black hole.}

\date{\today}

\begin{document}
\maketitle
\flushbottom

\section{Introduction}

Black holes are the most interesting objects in the universe.
Black hole has a event horizon, as a boundary where nothing can escape
include the light so that they are called as black \cite{luminet}. But
beyond that line, particles can escape. Black hole always try to pull the
surrounding matter which is known as accretion \cite{cunha,Cunha:2018acu}.
During the accretion of black holes, they release large amounts of energy
into their large-scale environments, so that black holes may play a
prominent role in the processes that control galaxy formation \cite%
{Falcke:1999pj,Tremblay:2016ijg}. Moreover, this accreting matter heats up
through viscous dissipation and radiate light in various frequencies. For
instance, the radio waves are one of them which can be detected through the
radio telescopes \cite{Shen:2005cw,Huang:2007us,Johannsen:2015mdd}. When the accretion happens
onto black hole, shining material pass through the event horizon, which
results in a dark area over a bright background: black hole shadow (BHS) 
\cite{Cunha:2018gql}. We can say that this shadow is actually an image of
the event horizon. The center of galaxies is a playground of a gigantic black holes. Because of
the gravitational lens effect, the background would have cast a shade larger
than its horizon size \cite{synge}. The size and shape of this shadow can be
calculated and visualized, respectively. 
 
In 1970s, Bardeen, Press and Teukolsky \cite{bardeen} and
then Chandrasekhar \cite{Chandra} computed that a BHS has a radius of $%
r_{shadow}=\sqrt{27}M=5.2M$ over the background light source (seen by an outside observer). As reported by
many numerical calculations, rotating black holes cast shadows of
approximately the same size as well \cite%
{Hioki:2009na,Johannsen:2010ru,Nedkova:2013msa,Amarilla:2013sj,Abdujabbarov:2012bn,Grenzebach:2014fha,Johannsen:2015hib,Giddings:2014ova,Atamurotov:2013sca,Wei:2013kza,Sakai:2014pga,Perlick:2015vta,Abdujabbarov:2015xqa,Tinchev:2013nba,Wang:2018eui,Tsukamoto:2017fxq,Amarilla:2010zq,Yumoto:2012kz,Takahashi:2005hy,Papnoi:2014aaa,Dexter:2012fh,Moffat:2015kva,Younsi:2016azx,Johannsen:2015qca,Zakharov:2014lqa,Cunha:2016bjh,Freivogel:2014lja,Cunha:2016bpi,Ohgami:2015nra,Zakharov:2011zz,Hennigar:2018hza,Pu:2016qak,Sharif:2016znp,Abdujabbarov:2016hnw,Xu:2018mkl,Gyulchev:2018fmd,Vetsov:2018mld,Hou:2018bar,Dokuchaev:2018kzk,Mizuno:2018lxz,Perlick:2018iye,Stuchlik:2018qyz,Shaikh:2018lcc,Eiroa:2017uuq,Mars:2017jkk,Wang:2017hjl,Singh:2017vfr,Mureika:2016efo,Huang:2016qnl,Ghasemi-Nodehi:2016wao,Tsukamoto:2014tja,Vincent:2016sjq}%
. On the other hand, null geodesics method and gravitational lensing are
also useful tools to gain information about the black holes \cite%
{Bartelmann:1999yn,Bozza:2009yw}. Strong and weak gravitational lensing by
black holes, wormholes or other exotic objects have been investigated by
several authors \cite%
{Virbhadra:2007kw,Zschocke:2011mm,Stefanov:2010xz,Manna:2018jxb,Ono:2018ybw}.

\begin{figure}[ht]
\begin{center}
\includegraphics[scale=0.1]{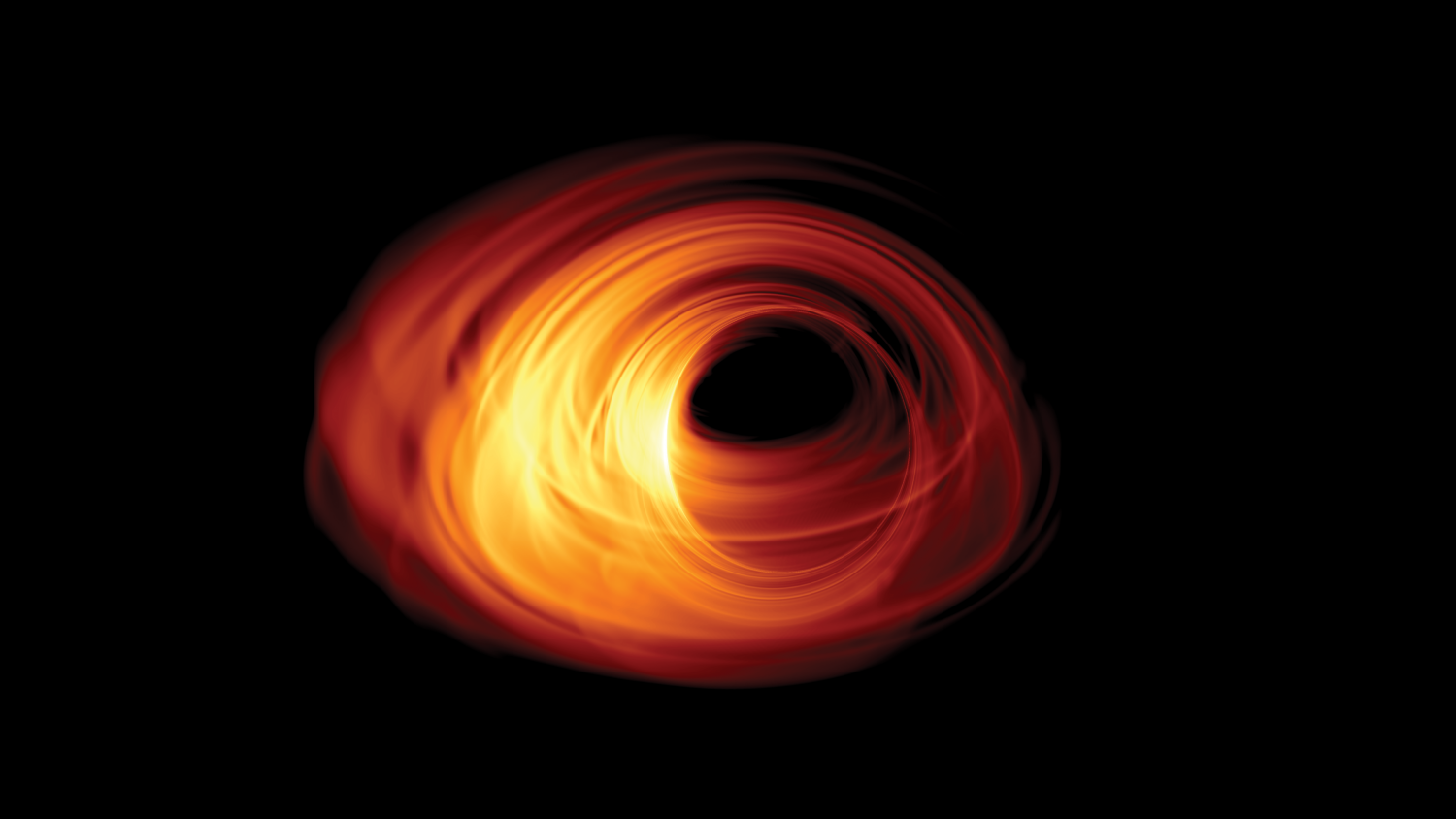}
\end{center}
\caption{Simulated image of an accreting black hole with shadow \protect\cite%
{sim}. }
\label{accreting}
\end{figure}

The main purpose of studying the gravitational lensing is to detect the
black holes in the universe, which are believed that they are mostly located
at the center of the galaxies. The strong gravitational lensing helps us to
find the position, magnification, and time delays of the images by black
holes. Moreover, in "weak lensing" the effect is much weaker but can still
be detected statistically \cite{Hu:2000ee}. To do so, Gibbons and Werner
generated a new technique to calculate the deflection angle of light rays
within the asymptotic source and receiver \cite{Gibbons:2008rj}. Their
method is based on the famous Gauss-Bonnet theorem (GBT), which solves the
integral in an infinite domain bounded by the light ray. After locating the
source and receiver to the asymptotic Minkowski regions, the deflection
angle of optical metric of a static black hole was computed \cite%
{Gibbons:2008hb}. Then Werner extended to the stationary black holes by
employing the Finsler-Randers type optical geometry with Naz{\i }m's
osculating Riemannian manifold \cite{Werner:2012rc}. In the sequel, this
topic has been thoroughly studied by several authors for different types of
spacetimes \cite%
{Gibbons:2015qja,Gibbons:2011rh,Gibbons:2008zi,Sakalli:2017ewb,Jusufi:2017lsl,Crisnejo:2018uyn,Jusufi:2017hed,Ovgun:2018ran,Jusufi:2017vta,Ovgun:2018prw,Jusufi:2017mav,Ovgun:2019prw,Jusufi:2017xnr,Cvetic:2016bxi,Jusufi:2017uhh,Ovgun:2018fnk,Jusufi:2018jof}%
. Recently, A. Ishihara et al. have shown that for the spherical symmetric static
objects, it is possible to find the deflection angle by considering the
finite-distance corrections instead of using the asymptotic receiver and
source \cite{Ishihara:2016vdc,Ishihara:2016sfv}. Then, T. Ono et al. have extended the method for the axisymmetric spacetimes \cite%
{Ishihara:2016vdc,Ishihara:2016sfv,Arakida:2017hrm}. The expectation of the
detection of the Milky Way's central supermassive black hole (Sagittarius A*
is the site of that black hole) by the Event Horizon Telescope (EHT), which is
simulated in Fig. \ref{accreting} increases the impact on the studies about
the BHS. EHT tries to observe the extremely hot gas around the
event horizon of the black hole \cite{dole}.

It is believed that there would exist formation (and creation ) mechanism of gravitomagnetic charge in the gravitational
interaction, just as some prevalent theories  \cite{iz1} that provide the theoretical mechanism of existence of magnetic monopole in various gauge interactions. Magnetic monopole in electrodynamics and gauge field theory has been extensively discussed and sought after for decades, and the existence of the 't Hooft-Polyakov monopole solution has spurred new interest of both theorists and experimentalists \cite{iz2,iz3}. The gravitomagnetic charge is the proposed gravitational analogue of Dirac's magnetic monopole \cite{iz4}. However, if it is indeed present in universe, it will also lead to significant consequences in astrophysics and cosmology. On the other hand, as it is well-known, dyon is a pole possessing both electric and magnetic charges. In the "weak field limit", Einstein's equations reduce to a form remarkably like Maxwell's equations of electromagnetism. Terms appear that are analogous to the electric field caused by charges and the magnetic field produced by the flow of charge. The "electric terms" correspond simply to the gravity that keeps our feet on the ground. The "magnetic terms" are wholly unfamiliar; we do not see them in everyday life. The KNK spacetime \cite{kasuya} is nothing but a rotating dyon (a hypothetical particle in 4-dimensional theories with both electric and magnetic charges) black hole. The best place to measure gravitomagnetism is in Earth orbit. Just as a spinning ball of electric charge produces a well-defined magnetic field, a spinning mass such as Earth is expected to produce a well-defined gravitomagnetic field. On the other hand, recent studies of gauge theories have given promising results about the existence of the monopole \cite{Chanyal:2017xqf,Mavromatos:2016ykh}. Besides, there is a renewed interest for the cosmological constant since it can be one of the theoretical models to explain the inflationary scenario of the early universe. In this scenario, the universe undergoes a stage which is geometrically described by hot de-Sitter(dS) spacetime which is related with KNK spacetime \cite{PLBds}. In
addition to this, it was shown that primordial universe can be described by the KNK spacetime \cite{Comtet:1979kq}. In the present study, we consider the KNK spacetime and its shadow cast. To this end, we employ the GBT and analyze the weak gravitational
lensing by the KNK black hole.

This paper is organized as follows. Section II briefly describes the KNK
black hole and serves its physical features. We compute the deflection angle by
following the GBT of \cite{Gibbons:2008rj} in Sec. III. Section IV is devoted to the computation of the
shadows of the KNK black holes by manipulating the rotation parameter and
the dyon charge. Conclusions are presented in Sec. V.

\section{KNK spacetime}

The metric of the KNK black hole in the Boyer-Lindquist coordinates is given
by \cite{kasuya} 
\begin{equation}
ds^{2}=-\frac{\Delta }{\Sigma}(dt-a\sin ^{2}\theta \ d\phi )^{2}+\frac{%
\Sigma }{\Delta }\ dr^{2}+\Sigma \ d\theta ^{2}+\frac{\sin ^{2}\theta }{%
\Sigma }[(r^{2}+a^{2})d\phi -a\ dt]^{2}\ ,
\label{eq:metrica_Kerr-Newman-Kasuya}
\end{equation}%
where 
\begin{equation}
\Delta =r^{2}-2Mr+a^{2}+Q_{e}^{2}+Q_{m}^{2}\ ,
\label{eq:Delta_metrica_Kerr-Newman-Kasuya}
\end{equation}%
and 
\begin{equation}
\Sigma =r^{2}+a^{2}\cos ^{2}\theta \ .
\label{eq:rho_metrica_Kerr-Newman-Kasuya}
\end{equation}

Note that $M$ is the mass, $a=J/M$ represents the rotation parameter, which
is the angular momentum per unit mass, $Q_{e}$ and $Q_{m}$ denote electric
and magnetic charges, respectively. The spacetime of the KNK black hole
reduces to the Kerr-Newman black hole when $Q_{m}=O$, the Kerr black hole if 
$Q_{e}=Q_{m}=0$, the Reissner-Nordstr\"om black hole for $Q_{m}=a=0$
and the Schwarzschild black hole if $a=Q_{e}=Q_{m}=0$. Meanwhile, the KNK spacetime is one of the members of Pleba$\acute{\textmd{n}}$ski-Demia$\acute{\textmd{n}}$ski family of black hole solutions \cite{iz6}. 

Event horizon of the KNK black hole is obtained by using the following
equation 
\begin{equation}
\Delta =(r-r_{+})(r-r_{-})=0\ ,  \label{eq:superficie_hor_Kerr-Newman-Kasuya}
\end{equation}%
whose solutions are 
\begin{equation}
r_{+}=M+[M^{2}-(a^{2}+Q_{e}^{2}+Q_{m}^{2})]^{1/2}\ ,
\label{eq:sol_padrao_Kerr-Newman-Kasuya_1}
\end{equation}%
and 
\begin{equation}
r_{-}=M-[M^{2}-(a^{2}+Q_{e}^{2}+Q_{m}^{2})]^{1/2}\ .
\label{eq:sol_padrao_Kerr-Newman-Kasuya_2}
\end{equation}%
The surface gravity \cite{Wald} of the KNK black hole can be obtained as
follows: 
\begin{equation}
\kappa _{+}\equiv \frac{1}{2}\frac{1}{r_{+}^{2}+a^{2}}\left. \frac{d\Delta }{%
dr}\right\vert _{r=r_{+}}=\frac{1}{2}\frac{r_{+}-r_{-}}{r_{+}^{2}+a^{2}}\ .
\label{eq:acel_grav_ext_Kerr-Newman-Kasuya}
\end{equation}

Thus, the Hawking temperature \cite{Wald}\ of the KNK black hole becomes 
\begin{equation}
T_{+}=\frac{\kappa _{+}}{2\pi }=\frac{1}{4\pi }\frac{r_{+}-r_{-}}{%
r_{+}^{2}+a^{2}}.  \label{eq:temp_Hawking_Kerr-Newman-Kasuya}
\end{equation}%
The surface area of the horizon is given by 
\begin{equation}
\mathcal{A}_{+}=\left. \int \int \sqrt{-g}\ d\theta \ d\phi \right\vert
_{r=r_{+}}=4\pi (r_{+}^{2}+a^{2})\,,  \label{eq:area_Kerr-Newman-Kasuya}
\end{equation}%
where $g\equiv \mbox{det}(g_{\sigma \tau })=-r^{4}\sin ^{2}\theta $. 

The entropy \cite{Wald} of the KNK black hole at the event horizon reads 
\begin{equation}
S_{+}=\frac{\mathcal{A}_{+}}{4}=\pi (r_{+}^{2}+a^{2}).  \label{eq:entropia_Kerr-Newman-Kasuya}
\end{equation}%
The angular velocity of KNK black hole is given by 
\begin{equation}
\Omega _{+}=\left. -\frac{g_{t\phi}}{g_{\phi \phi}}\right\vert _{r=r_{+}}=\frac{a}{%
(r_{+}^{2}+a^{2})}\ .  \label{eq:vel_ang_Kerr-Newman-Kasuya}
\end{equation}%
Furthermore, the electric potential with magnetic and electric charges is as
follows 
\begin{equation}
\Phi _{+}=\frac{Q_{e}r_{+}-\xi Q_{m}a}{r_{+}^{2}+a^{2}}\ .
\label{eq:pot_ele_Kerr-Newman-Kasuya}
\end{equation}%
Note that $\xi =\pm 1$ stands for the two gauges. The first law of the thermodynamics is satisfied via the
following expression: 
\begin{equation}
dM=T_{+}\ dS_{+}+\Omega _{+}\ dJ+\Phi _{+}\ dQ_{e}\ .
\label{eq:1_lei_termo_Kerr-Newman-Kasuya}
\end{equation}

\section{Deflection angle of light by KNK spacetime}

In this section, by using the GBT we shall study the deflection angle for
the KNK black hole. We use the null condition $ds^{2}=0$ and solve the KNK
spacetime for $dt$ as follows: 
\begin{equation}
dt=\sqrt{\gamma _{ij}dx^{i}dx^{j}}+\beta _{i}dx^{i}.  \label{opt}
\end{equation}%
Note that $\gamma _{ij}$ goes as ($i,j=1,2,3$). Then, we obtain the
components of the optical metric of KNK spacetime in terms of $\gamma _{ij}$
and $\beta _{i}$: 
\begin{align}
d\ell ^{2}=\gamma _{ij}dx^{i}dx^{j}=& \frac{\Sigma ^{2}}{\Delta (\Sigma -2Mr)%
}dr^{2}+\frac{\Sigma ^{2}}{(\Sigma -2Mr)}d\theta ^{2}+\left( r^{2}+a^{2}+%
\frac{2a^{2}Mr\sin ^{2}\theta }{(\Sigma -2Mr)}\right) \frac{\Sigma \sin
^{2}\theta }{(\Sigma -2Mr)}d\phi ^{2},  \label{gamma-KNK} \\
\beta _{i}dx^{i}=& -\frac{2aMr\sin ^{2}\theta }{(\Sigma -2Mr)}d\phi .
\label{beta-Kerr}
\end{align}

The spatial metric $\gamma _{ij}$ stands for an arc-length ($\ell $) along
the light ray. At the equatorial plane ($\theta =\pi /2$), we have

\begin{equation}
d\ell ^{2}=\frac{\Sigma ^{2}}{\Delta (\Sigma -2Mr)}dr^{2}+\left( r^{2}+a^{2}+%
\frac{2a^{2}Mr}{(\Sigma -2Mr)}\right) \frac{\Sigma }{(\Sigma -2Mr)}d\phi
^{2},
\end{equation}%
and 
\begin{equation}
\beta _{i}dx^{i}=-\omega d\phi ,
\end{equation}%
where $\omega =\frac{2aMr}{(\Sigma -2Mr)}d\phi $. In practice, we define the
deflection angle ($\hat{\alpha}$) with the angles $\Psi _{R}$, $\Psi _{S}$
and $\phi _{RS}$ which correspond to receiver, source, and coordinate,
respectively. Namely, we have 
\begin{equation}
\hat{\alpha}\equiv \Psi _{R}-\Psi _{S}+\phi _{RS}.  \label{alpha-axial}
\end{equation}%
Locating the position of the receiver and source to suitable locations
where the endpoints of light rays lie in the Euclidean space, the GBT {\cite%
{Ono:2017pie}} admits the following expression for the deflection angle:

\begin{equation}
\hat{\alpha}=-\iint_{_{R}^{\infty }\square _{S}^{\infty
}}KdS+\int_{S}^{R}\kappa _{g}d\ell .  \label{GB-axial}
\end{equation}%
We should note that $_{R}^{\infty }\square _{S}^{\infty }$ is embedded
quadrilateral. Moreover, $\kappa _{g}$ is for the geodesic curvature and $%
d\ell $ is for an arc length.  For evaluating the first integral of Eq. (\ref%
{GB-axial}), one should first calculate the Gaussian optical curvature
(related with 2-dimensional Riemann tensor) in the weak field approximation:

\begin{equation}
K=\frac{R_{r\phi r\phi }}{\det \gamma }=-2\,{\frac{M}{{r}^{3}}}-3\,{\frac{%
\left( -r+2\,M\right) {Q_{m}}^{2}}{{r}^{5}}}-3\,{\frac{\left( -r+2\,M\right) 
{Q_{e}}^{2}}{{r}^{5}}}.
\end{equation}%

Then, we compute the following geodesic curvature 
\cite{Ono:2017pie}:

\begin{equation}
\kappa _{g}=-\sqrt{\frac{1}{\gamma \gamma ^{\theta \theta }}}\beta _{\phi
,r}, \label{kappa}
\end{equation}
where $\gamma$ is the $det \gamma$.

For the KNK black hole, the geodesic curvature $\kappa _{g}$ is computed as 
\begin{equation}
\kappa _{g}=-\frac{2aM}{r^{3}}.
\end{equation}

We now examine the net contribution of the geodesic part: 
\begin{equation}
\int_{S}^{R}\kappa _{g}d\ell =\int_{~R}^{S}\frac{2aM}{r^{3}}d\ell =\frac{4aM%
}{b^{2}},  \label{intkappa}
\end{equation}
where $b$ is the impact parameter.
To visualize the boundary of the integration domain, we define the angle by
using the outgoing radial direction of the light rays:

\begin{equation}
\sin \Psi =\frac{\sqrt{\beta ^{2}+r^{2}}b}{r^{2}}-\frac{\beta ^{2}a}{r^{4}%
\sqrt{\beta ^{2}+r^{2}}}-\frac{bM}{r\sqrt{\beta ^{2}+r^{2}}},  \label{u}
\end{equation}

and the solution for the photon orbit is found to be

\begin{equation}
u=\frac{\sin \phi }{b}+\frac{M(1+\cos ^{2}\phi )}{b^{2}}-\frac{2aM}{b^{3}}.
\end{equation}

Afterward, we calculate the integral of the Gaussian curvature of the
optical metric of the KNK black hole \cite{Gibbons:2008hb}

\begin{equation}
-\iint_{_{R}^{\infty }\square _{S}^{\infty }}KdS=\int_{0}^{\pi }\int_{\infty
}^{u}Kdrd\phi =\frac{4M}{b}-\frac{3Q_{e}^{2}}{4b^{2}}-\frac{3Q_{m}^{2}}{%
4b^{2}}.  \label{intK}
\end{equation}%
In sequel, we combine the solutions of the Gaussian optical curvature
integral and geodesic curvature integral to obtain the total deflection
angle of light on the equatorial plane of the KNK black hole: 
\begin{equation}
\hat{\alpha}=\frac{4M}{b}-\frac{3Q_{e}^{2}}{4b^{2}}-\frac{3Q_{m}^{2}}{4b^{2}}%
\pm \frac{4aM}{b^{2}}.  \label{Dangle}
\end{equation}%
We note that the positive sign stands for the retrograde and negative sign
is for the prograde case of the photon orbit. The deflection angle of the
KNK black hole is agreed with the Kerr case with the limit of $Q_{e}=Q_{m}=0$ \cite{kerr}
and the non-rotating dyon black hole if $a=0$ \cite{Clem1}. Moreover, the deflection
angle of the charged black hole is recovered if $a=Q_{m}=0$ and the
deflection angle of the Schwarzschild black hole is obtained for $%
a=Q_{e}=Q_{m}=0$. The deflection angle is linearly decreased with the extra
magnetic charge compared to Kerr-Newman black hole.


\section{Shadows of KNK spacetime}

The KNK spacetime in Boyer-Lindquist coordinates $g_{\mu \nu }^{\mathrm{K}}$
is described by the following metric tensors: 
\begin{eqnarray}
g_{tt}^{\mathrm{KNK}} &=&-\left( 1-\frac{2Mr}{\Sigma }\right) ,  \nonumber \\
g_{t\phi }^{\mathrm{KNK}} &=&-\frac{2Mar\sin ^{2}\theta }{\Sigma }, 
\nonumber \\
g_{rr}^{\mathrm{KNK}} &=&\frac{\Sigma }{\Delta },  \nonumber \\
g_{\theta \theta }^{\mathrm{KNK}} &=&\Sigma ,  \nonumber \\
g_{\phi \phi }^{\mathrm{KNK}} &=&\left( r^{2}+a^{2}+\frac{2Ma^{2}r\sin
^{2}\theta }{\Sigma }\right) \sin ^{2}\theta ,  \label{kerr}
\end{eqnarray}

where 
\begin{eqnarray}
\Delta &\equiv& r^2-2Mr+a^2+Q_m^2+Q_e^2,  \nonumber \\
\Sigma &\equiv& r^2+a^2\cos^2 \theta.  \label{deltasigma}
\end{eqnarray}

The motion of the particle on the KNK is obtained by the following
Lagrangian: \cite{Chandra} 
\begin{equation}
\mathcal{L}=\frac{1}{2}g_{\nu \sigma }\dot{x}^{\nu }\dot{x}^{\sigma },
\label{r}
\end{equation}%
where $\dot{x}^{\nu }=u^{\nu }=dx^{\nu }/d\lambda $ in which $u^{\nu }$ stands for
four velocity of the particle with the affine parameter $\lambda $. Due to
the symmetry of black hole, conjugate momenta \ $p_{t}$ and $p_{\phi }$ are
conserved due to the metric-independent variables $t$ and $\phi $. Hence, the
energy $E$ and angular momentum $L$ are obtained as: 
\begin{equation}
E=p_{t}=\frac{\partial \mathcal{L}}{\partial \dot{t}}=g_{\phi t}\dot{\phi}%
+g_{tt}\dot{t},\quad L=-p_{\phi }=-\frac{\partial \mathcal{L}}{\partial \dot{%
\phi}}=-g_{\phi \phi }\dot{\phi}-g_{\phi t}\dot{t}.  \label{E}
\end{equation}%

and we derive 
\begin{eqnarray}
&&\Sigma \dot{t}=-a(aE\sin ^{2}\theta -L)+\frac{(r^{2}+a^{2})P(r)}{\Delta (r)%
}, \\
&&\Sigma \dot{\phi}=-\left( aE-\frac{L}{\sin ^{2}\theta }\right) +\frac{aP(r)%
}{\Delta (r)},
\end{eqnarray}%
where $P(r)\equiv E(r^{2}+a^{2})-aL$. To calculate the other geodesics
equations, we use the Hamilton-Jacobi equation: 
\[
\frac{\partial S}{\partial \lambda }=\frac{1}{2}g^{\nu \sigma }\frac{%
\partial S}{\partial x^{\nu }}\frac{\partial S}{\partial x^{\sigma }},
\]%
with the following ansatz: 
\[
S=\frac{1}{2}\mu ^{2}\lambda -Et+L\phi +S_{r}(r)+S_{\theta }(\theta ),
\]%
where $\mu$ is proportional to the rest mass of the particle.

For the KNK spacetime, the Hamilton-Jacobi equation yields: 
\[
\frac{1}{2}g^{tt}\frac{\partial S}{\partial x^{t}}\frac{\partial S}{\partial
x^{t}}+g^{\phi t}\frac{\partial S}{\partial x^{t}}\frac{\partial S}{\partial
x^{\phi }}+\frac{1}{2}g^{rr}\frac{\partial S}{\partial x^{r}}\frac{\partial S%
}{\partial x^{r}}+\frac{1}{2}g^{\theta \theta }\frac{\partial S}{\partial
x^{\theta }}\frac{\partial S}{\partial x^{\theta }}+\frac{1}{2}g^{\phi \phi }%
\frac{\partial S}{\partial x^{\phi }}\frac{\partial S}{\partial x^{\phi }}=-%
\frac{\partial S}{\partial \lambda }.
\]

Then, we solve it for $S_r$ and $S_\theta$ as follows \cite{Chandra}:

\begin{eqnarray}
&&\Sigma \frac{\partial S_{r}}{\partial r}=\pm \sqrt{R(r)},
\label{eq:Theta_motion} \\
&&\Sigma \frac{\partial S_{\theta }}{\partial \theta }=\pm \sqrt{\Theta
(\theta )},
\end{eqnarray}%
where 
\begin{eqnarray}
&&R(r)\equiv P(r)^{2}-\Delta (r)\left[ (L-aE)^{2}+\mathcal{Q}\right] , \\
&&\Theta (\theta )\equiv \mathcal{Q}+\cos ^{2}\theta \left( a^{2}E^{2}-\frac{%
L^{2}}{\sin ^{2}\theta }\right) ,
\end{eqnarray}%
and $\mathcal{Q}$ is the Carter constant defined by $\mathcal{Q}\equiv 
\mathcal{K}-(L-aE)^{2}$ where $\mathcal{K}$ is a constant of motion. $R(r)$ and $\Theta (\theta )$ should be
non-negative for the photon motion. Two impact parameters $\eta $ and $\xi $
are introduced in terms of energy $E$, angular momentum $L$ and Carter
constant $\mathcal{Q}$ as \cite{Chandra} 
\begin{equation}
\xi \equiv \frac{L}{E},\quad \quad \eta \equiv \frac{\mathcal{Q}}{E^{2}},
\end{equation}%
For photon case, Eq.~(\ref{r}) can be rewritten in terms of dimensionless
quantities $\eta $ and $\xi $: 
\begin{equation}
R(r)=\frac{1}{{E}^{2}}\left[ (r^{2}+a^{2})-a{\xi }\right] ^{2}-\Delta \left[
(a-{\xi })^{2}+{\eta }\right] .
\end{equation}

Note that $R$ and $\Theta$ act as effective potentials for moving particle
in $r$ and $\theta$ directions, respectively. Equation $S_r$ can be represented as 
\begin{equation}
( \frac{\partial S_r}{\partial r})^2 + V_{eff} = 0,
\end{equation}
where $V_{eff}$ is the effective potential:
\begin{equation}
V_{eff}=\frac{1}{{{\Sigma }^{2}}}\left[ (r^{2}+a^{2})-a{\xi }\right]
^{2}-\Delta \left[ (a-{\xi })^{2}+{\eta }\right] .  \label{vef}
\end{equation}%
We can achieve the most critical and unstable circular orbit by maximizing
the effective potential, which should satisfy the following conditions 
\begin{equation}
V_{eff}=\left. \frac{\partial V_{eff}}{\partial r}\right\vert
_{r=r_{0}}=0\;\;\;\;\mbox{or}\;\;\;R=\left. \frac{\partial R}{\partial r}%
\right\vert _{r=r_{0}}=0,  \label{vr}
\end{equation}%
where $r=r_{0}$ is the radius of the unstable circular null orbit. We assume
that the photons and the observer are located at the infinity ($\mu =0$) and
photons come near the equatorial plane $(\theta =\frac{\pi }{2})$. We solve
the Eq. (\ref{vr}) and obtain the following celestial coordinates of the
image: 
\begin{eqnarray}
\xi  &=&{\frac{r^{2}-r\Delta -a^{2}}{a(r-1)}},  \label{17} \\
\eta  &=&{\frac{r^{3}[4\Delta -r(r-1)^{2}]}{a^{2}(r-1)^{2}}}.  \label{17aa}
\end{eqnarray}%
Note that observer at spatial infinity can observe the celestial coordinates
of the image and determine the contour of the BHS. We briefly review the
expressions of the images of photon rings around black holes. There are
three independent constants of motion in the KNK metric. The inertial reference
frame of an observer with the basis vectors $\{e_{\hat{t}},e_{\hat{r}},e_{%
\hat{\theta}},e_{\hat{\phi}}\}$ and coordinate basis of the metric $%
\{e_{t},e_{r},e_{\theta },e_{\phi }\}$ \cite{Johannsen:2010ru}: 
\begin{eqnarray}
e_{\hat{t}} &=&\zeta e_{t}+\gamma e_{\phi },  \nonumber \\
e_{\hat{r}} &=&\frac{1}{\sqrt{g_{rr}}}e_{r},  \nonumber \\
e_{\hat{\theta}} &=&\frac{1}{\sqrt{g_{\theta \theta }}}e_{\theta }, 
\nonumber \\
e_{\hat{\phi}} &=&\frac{1}{\sqrt{g_{\phi \phi }}}e_{\phi },
\label{eq:localframe}
\end{eqnarray}%
which are defined, at a large distance from the KNK black hole, with the
relation between local observer's basis vector and coordinate basis of
metric: $e_{\hat{\alpha}}=e_{~\hat{\alpha}}^{\mu }e_{\mu }$ with $e_{~\hat{%
\alpha}}^{\mu }e_{~\hat{\beta}}^{\nu }g_{\mu \nu }=\eta _{{\hat{\alpha}}{%
\hat{\beta}}}$ for the Minkowski metric $\eta _{{\hat{\alpha}}{\hat{\beta}}}=%
\mathrm{diag}(-1,1,1,1)$. From the orthonormal property of $\zeta $ and $%
\gamma $, one obtains the following equations \cite%
{bardeen,Johannsen:2015qca}: 
\begin{eqnarray}
\zeta  &=&\sqrt{\frac{g_{\phi \phi }}{g_{t\phi }^{2}-g_{tt}g_{\phi \phi }}},
\nonumber \\
\gamma  &=&-\frac{g_{t\phi }}{g_{\phi \phi }}\sqrt{\frac{g_{\phi \phi }}{%
g_{t\phi }^{2}-g_{tt}g_{\phi \phi }}}.  \label{eq:zetagamma}
\end{eqnarray}%
Then, the energy and angular momentum are obtained as: 
\begin{eqnarray}
p^{\hat{t}} &=&\zeta E-\gamma L, \\
p^{\hat{\phi}} &=&\frac{1}{\sqrt{g_{\phi \phi }}}L.
\end{eqnarray}%
On the other hand, the quantities $\alpha $ and $\beta $ are known as the
impact parameters. They are the axes of the Cartesian coordinate of the
image plane of the observer located at a distance $r=r_{0}$ and inclination
angle $\theta =i$ between the rotation axis of KNK black hole and observer's
line of sight: 
\begin{eqnarray}
\alpha  &\equiv &-r_{0}\frac{p^{\hat{\phi}}}{p^{\hat{t}}}=-r_{0}\frac{\xi }{%
\sqrt{g_{\phi \phi }}\zeta \left( 1+\frac{g_{t\phi }}{g_{\phi \phi }}\xi
\right) }, \\
\beta  &\equiv &r_{0}\frac{p^{\hat{\theta}}}{p^{\hat{t}}}=r_{0}\frac{\pm 
\sqrt{\Theta (i)}}{\sqrt{g_{\theta \theta }}\zeta \left( 1+\frac{g_{t\phi }}{%
g_{\phi \phi }}\xi \right) },
\end{eqnarray}%
where 
\begin{eqnarray}
p^{\hat{\theta}} &=&\frac{p_{\theta }}{\sqrt{g_{\theta \theta }}}=\frac{\pm 
\sqrt{\Theta (i)}}{\sqrt{g_{\theta \theta }}}, \\
\Theta (i) &\equiv &\eta +a^{2}\cos ^{2}i-\xi ^{2}\cot ^{2}i.
\end{eqnarray}%
All the metric elements seen in the above expressions are evaluated at $%
r=r_{0}$ and $\theta =i$. In the limit $r_{0}\rightarrow \infty $, the shape
of the BHS for a far away observer can be determined by the following
celestial coordinates \cite{bardeen,Chandra}: 
\begin{eqnarray}
\alpha  &=&\lim_{r\rightarrow \infty }(-r^{2}\sin \theta \frac{d\phi }{dr}%
|_{\theta \rightarrow i}),  \label{18} \\
\beta  &=&\pm \lim_{r\rightarrow \infty }(r^{2}\frac{d\theta }{dr}|_{\theta
\rightarrow i}),  \label{19}
\end{eqnarray}

which yield 
\begin{eqnarray}
\alpha  &=&-\xi \csc i,  \label{18a} \\
\beta  &=&\pm \sqrt{\eta +a^{2}\cos ^{2}i-\xi ^{2}\cot ^{2}i}.  \label{19a}
\end{eqnarray}%
The celestial coordinates $\alpha $ and $\beta $ show the apparent
perpendicular distances of the image around the black hole. For the KNK
black hole, at $i=\pi /2$, the celestial coordinates become 
\begin{eqnarray}
\alpha  &=&-\left( {\frac{r^{2}-r\Delta -a^{2}}{a(r-1)}}\right),  \\
\beta  &=&\pm \sqrt{{\frac{r^{3}[4\Delta -r(r-1)^{2}]}{a^{2}(r-1)^{2}}}}.
\label{29}
\end{eqnarray}

\begin{figure}[ht]
\begin{center}
\includegraphics[scale=0.5]{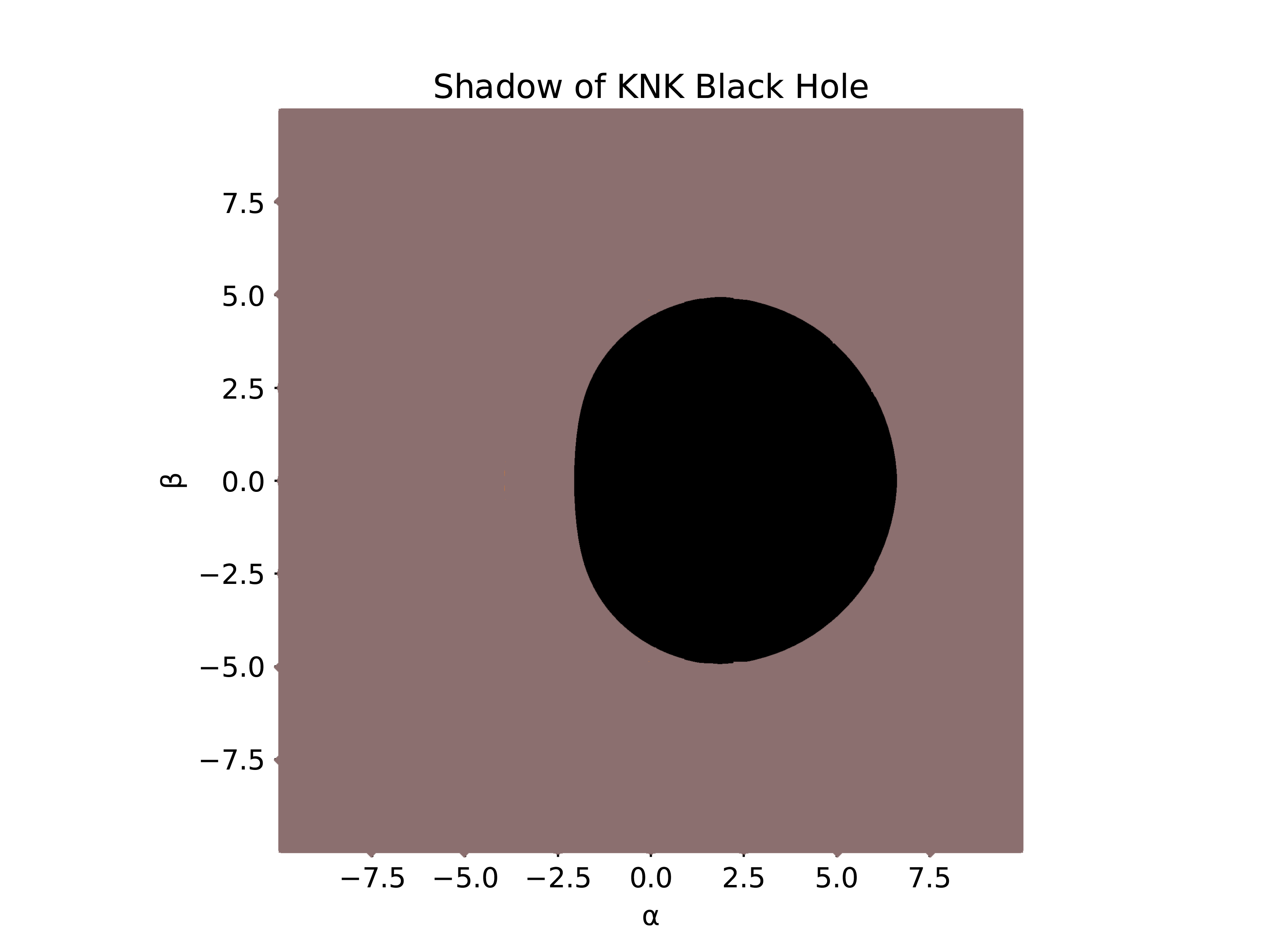}
\end{center}
\caption{Shadow of KNK black hole for $Q_e=0.1$, $Q_m=0.1$, $M = 1$, and
$a = 0.99$.}
\label{fig:shad}
\end{figure}

\begin{figure}[ht]
\includegraphics[scale=0.5]{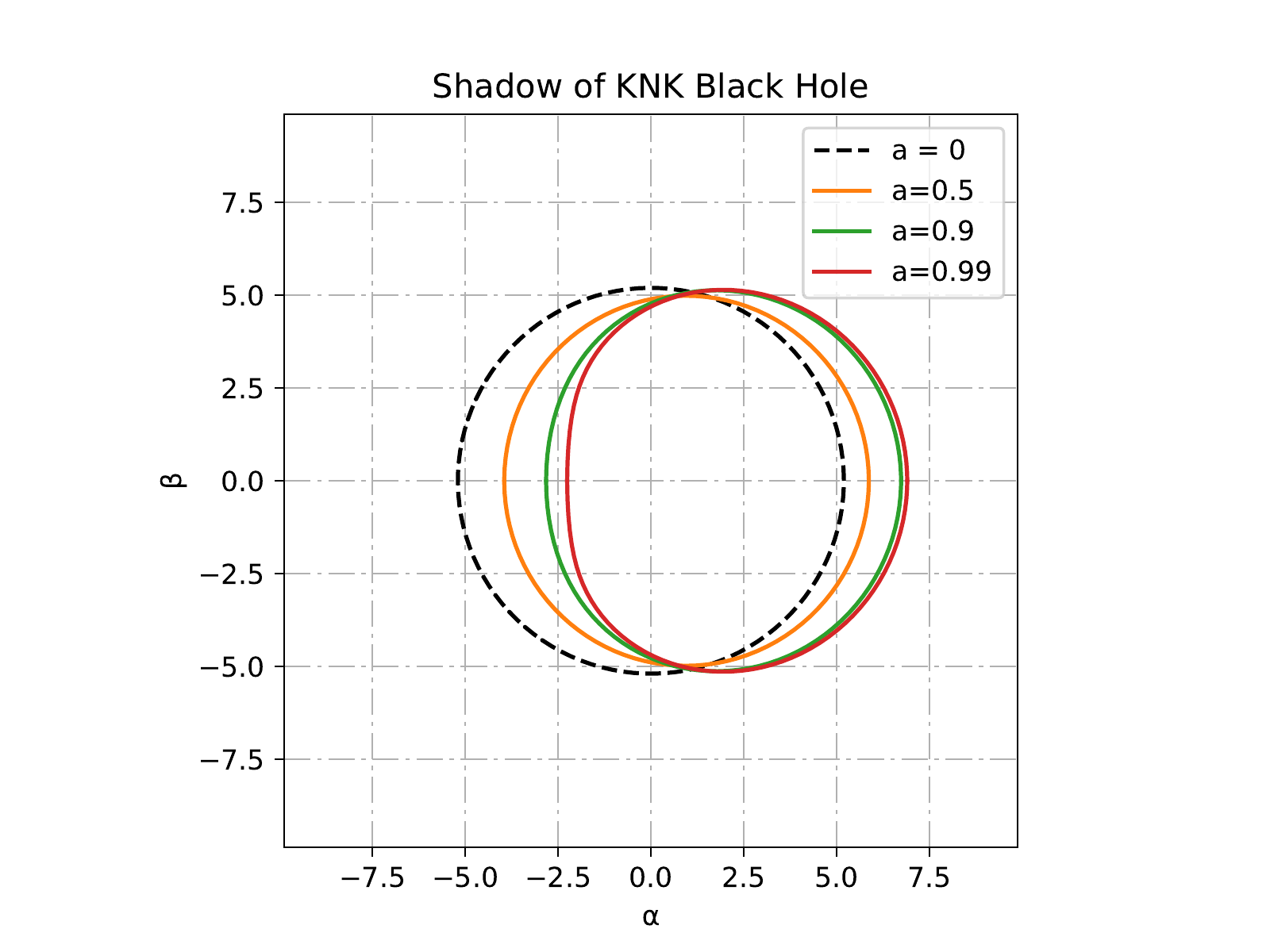} \includegraphics[scale=0.5]{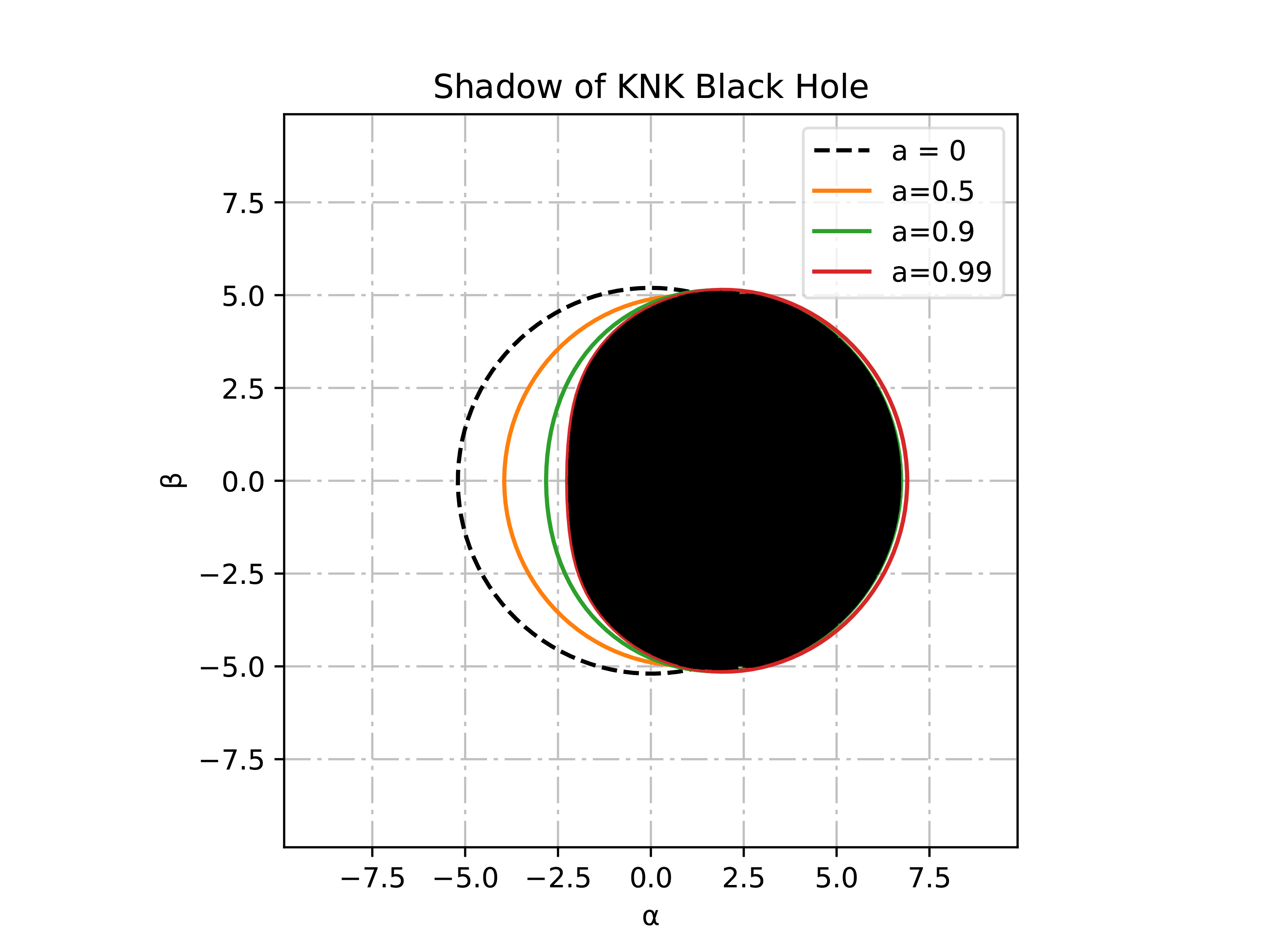} %
\includegraphics[scale=0.5]{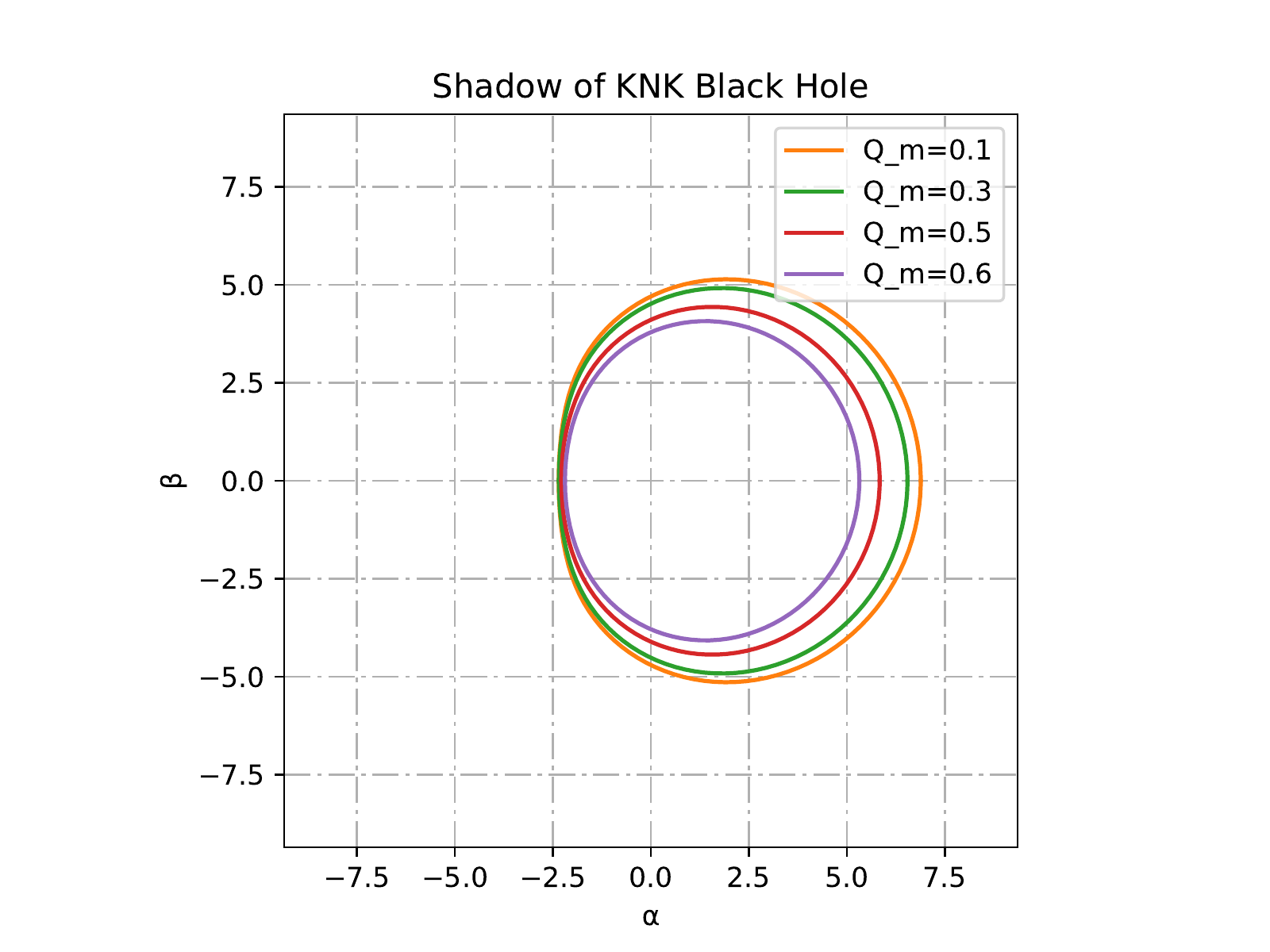} %
\includegraphics[scale=0.5]{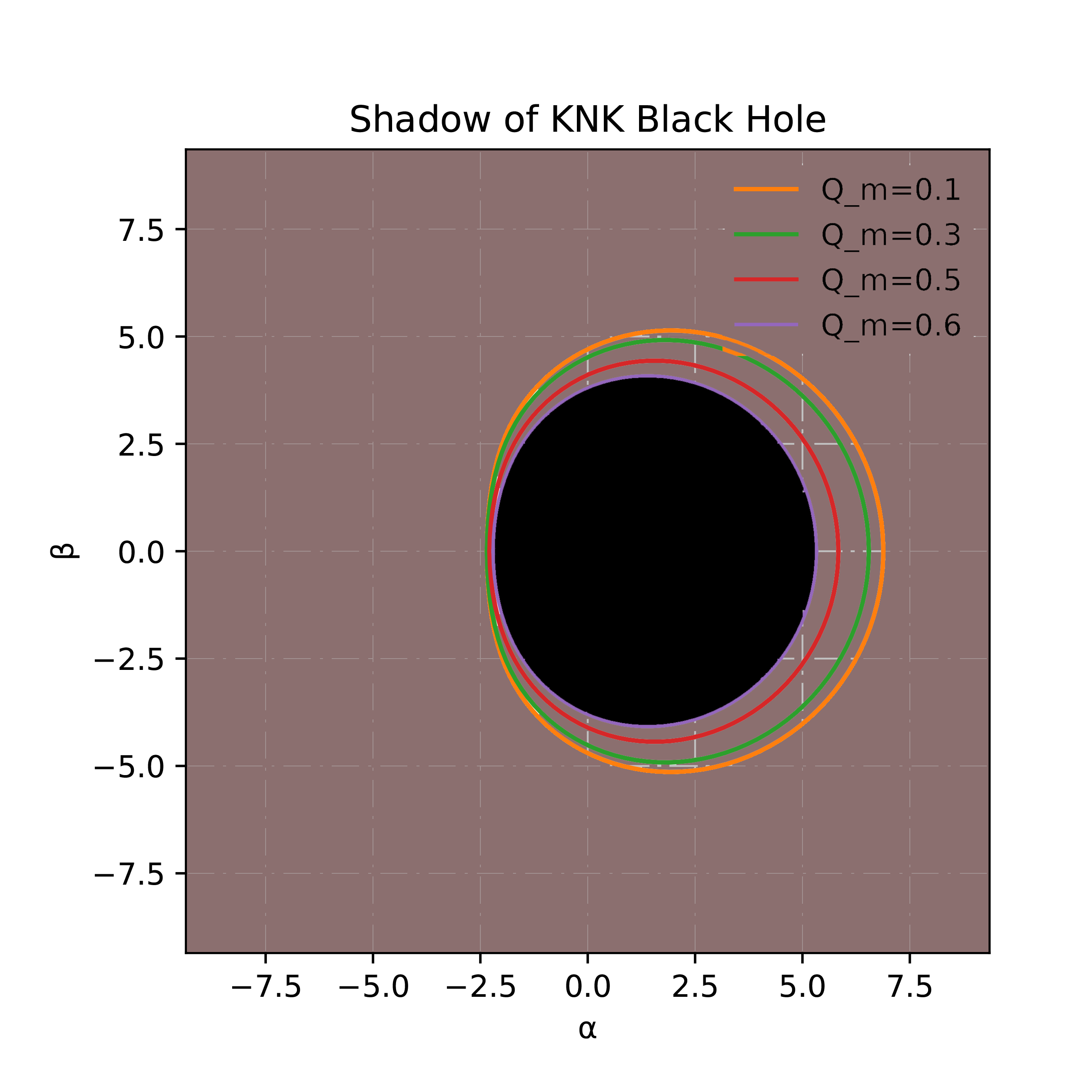}
\caption{Shadows of KNK black hole with different values of spin $a$ and
magnetic charge $Q_m$. The region bounded by each curve corresponds to the
black hole's shadow where the observer is at spatial infinity and in the equatorial
plane ($i = \protect\pi/2$). Left/Right side of the figures are prograde circular photon orbit and retrograde circular photon orbit, respectively. Top figures: $Q_e=0.1$, $Q_m = 0.1$ and $M = 1$. Bottom figures: $Q_e=0.1$, $M = 1$, and
$a = 0.9$.}
\label{fig:KNKa}
\end{figure}

\begin{figure}[ht]
\begin{center}
\includegraphics[scale=0.5]{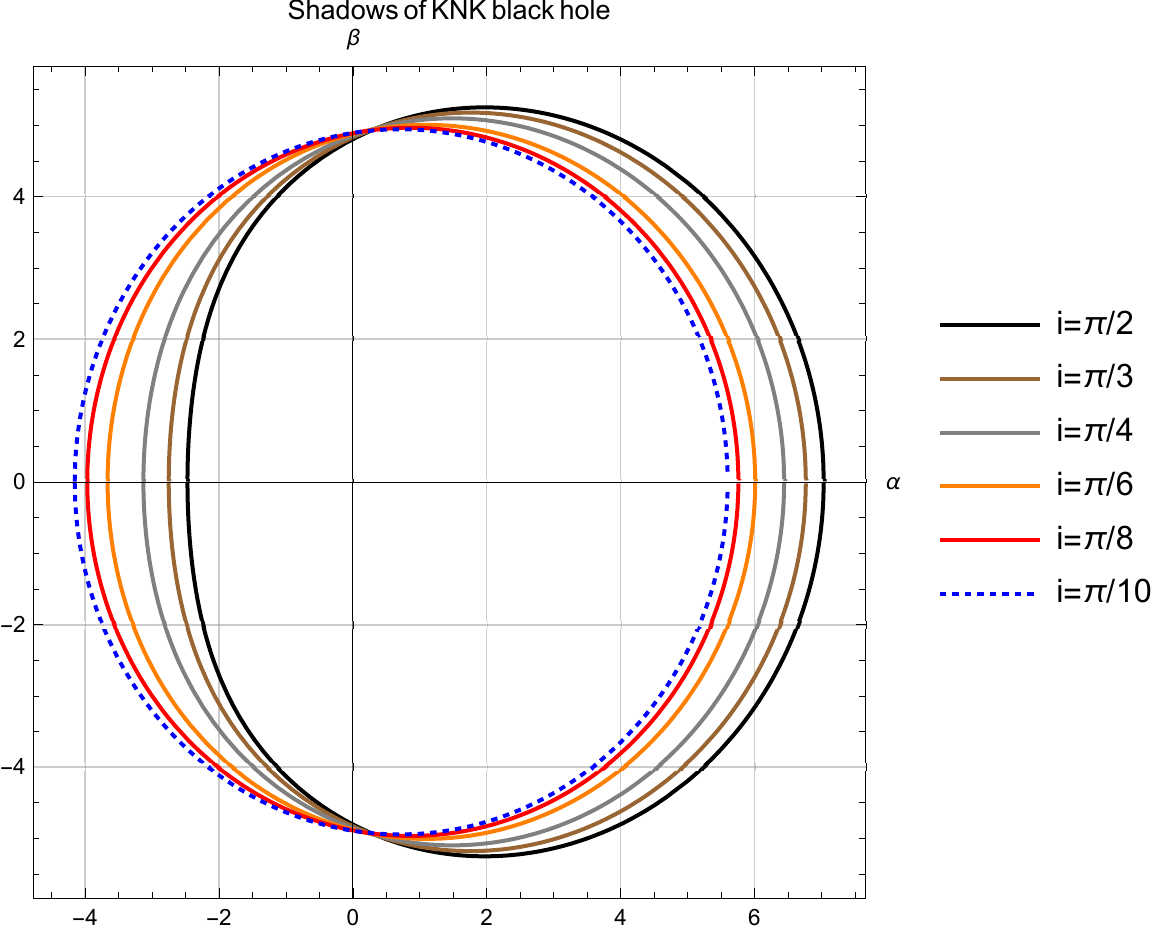}
\end{center}
\caption{ Photon rings are shown at inclination angles $i$ for $Q_e=0.1$, $Q_m=0.1$, $M = 1$, and
$a = 0.97$.} 
\label{fig:KNKi}
\end{figure}

Figures \ref{fig:shad} and \ref{fig:KNKa} show the plots of the shadows of the
KNK black hole with different values of spin $a$ and magnetic charge $Q_{m}$%
. Top figures of Fig. \ref{fig:KNKa} show the Schwarzschild case with dashed lines to compare with KNK black hole. For a fixed $a$ value, the presence of a charge $Q_m$ leads to a smaller shadow than in the case of Kerr geometry, and a value of $Q_m$ gives a more distorted shadow compared to Kerr-Newmann black hole. The shadows of KNK black hole with the different inclination
angles are also plotted in Fig. \ref{fig:KNKi}.

The EHT tries to understand the "event horizon" of two galactic center black holes such as Sgr A* black hole in the Milky Way galaxy and supermassive black hole in galaxy M87 which is about 1500 times more massive and 2000 times farther away than Sgr A*. However, the size of the black hole's event horizon in M87 is smaller than the event horizon of the Sgr A* black hole, on the other hand,  it is large enough for the EHT to resolve. For example: M87 black hole is about 22 micro-arc-sec as compared to the 53 micro-arc-sec of Sgr A* black hole so that the EHT requires strong angular resolution to match the small angular size of these black holes. The angular size of the shadow can be calculated
\begin{equation} \theta_s = R_s M/D_o,\end{equation}
where  $R_s$ is the angular radius, $D_o$ is the distance from the observer to the black hole and $M$ is the mass of the black hole. For the supermassive black hole Sgr A*, we have $M = 4.3 \times 10^6 M_{\odot}$ and $D_o = 8.3 kpc$ \cite{Gillessen:2008qv}; then, for the fixed parameters $Q_e=0.1$ and
$a = 0.9$, we obtain:

\begin{center}
\begin{tabular}{|c|c|c|c|c|c|c|c|c|}
\hline
$Q_m$ & $0.1$ & $0.3$ & $0.5$ & $0.6$ \\
\hline
$\theta _s (\mu \mathrm{as})$ & $26.77$ & $25.91$ & $24.29$ & $22.45$ \\
\hline
\end{tabular}
\end{center}
Here, the table shows that resolutions are needed to gain information from observations of the shadow of the Sgr A* black hole.

\section{Conclusion}

Alongside the recent successes of the LIGO detector and gravitational wave
astronomy \cite{LIGO}, the EHT has a potential outcome
to answer many questions at strong gravitational field regime in general
relativity by using the millimeter wavelength radio astronomy \cite{EHT} and
observe the BHS which is embedded on the image of hot gas.

First, we have studied the weak gravitational lensing by using the GBT for
the KNK optical spacetime, where the optical geometry gives a more
geometrical view on weak gravitational lensing than the quasi-Newtonian
lensing method. Remarkably, the method of GBT is easier than the null
geodesics method since we only use the optical geometry dealing with spatial
light rays. As a consequence of the GBT, the result obtained has a global
effect. When the locations of receiver and source are at null infinity, the
deflection angle in the weak field approximation has been found as follows
Eq. (\ref{Dangle}): 
\begin{equation}
\hat{\alpha}=\frac{4M}{b}-\frac{3Q_{e}^{2}}{4b^{2}}-\frac{3Q_{m}^{2}}{4b^{2}}%
\pm \frac{4aM}{b^{2}},
\end{equation}%
which has the Kerr limit at $Q_{e}=Q_{m}=0$ and the non-rotating dyon black
hole when $a=0$. Furthermore, the deflection angle of the charged black hole
is recovered if $a=Q_{m}=0$ and the deflection angle of the Schwarzschild
black hole is found for $a=Q_{e}=Q_{m}=0$. It is worth noting that the
deflection angle is linearly decreased with the extra magnetic charge
compared to Kerr-Newman black hole.

Secondly, we have investigated the shadow cast of the KNK black hole. With the
choice of this black hole, we have analyzed how $a,$ $Q_{e},$ and $Q_{m}$
parameters affect the image of the shadow. From the numerical plots, we have
shown that the magnetic charge or magnetic monopole $Q_{m}$ dramatically
decreases the size of the BHS cast. On the other hand, as seen in Fig. \ref%
{fig:KNKi}, the shadow cast of KNK black hole decreases with decreasing
inclination angle $i$.

In the near future, we are very hopeful that the EHT
will provide the event horizon visualization within black hole shadow. If
this observation occurs, we expect to see some of the experimental results
of this theoretical work.

\acknowledgments
This work is supported by Comisi{\'o}n Nacional de
Ciencias y Tecnolog{\'i}a of Chile (CONICYT) through FONDECYT Grant N{$%
\mathrm{o}$} 3170035 (A. {\"O}.). A. \"{O}. is grateful to Institute for
Advanced Study, Princeton for hospitality.

\end{document}